\documentclass[usenatbib,usegraphicx]{mn2e}
\usepackage{natbib}
\bibpunct{(}{)}{;}{a}{}{,}
\usepackage{graphicx}
\usepackage{psfig}
\usepackage{times}
\usepackage{amsmath}
\usepackage{amssymb}

\title{Dynamics of magnetized relativistic tori oscillating around  black holes}

\author[Montero et al]{P J Montero$^1$, O Zanotti$^2$, J A Font$^1$
  and L Rezzolla$^{3,4}$ \\
  $^1$ Departamento de Astronom\'{i}a y  Astrof\'{i}sica, Universidad de
  Valencia, Dr. Moliner 50, 46100 Burjassot (Valencia), Spain \\
  $^2$ Dipartimento di Astronomia e Scienza dello Spazio,
  Universita di Firenze, Firenze, Italy \\
 $^3$ Max-Planck-Institut f\"{u}r Gravitationsphysik, Albert-Einstein-Institut, Golm, Germany \\
  $^4$ Department of Physics, Louisiana State University, Baton
  Rouge, LA 70803, USA}	
  
\date{}
\pagerange{\pageref{firstpage}--\pageref{lastpage}}
\pubyear{2007}

\begin{document}

\maketitle
\label{firstpage}

\begin{abstract}
  We present a numerical study of the dynamics of magnetized,
  relativistic, non-self-gravitating, axisymmetric tori orbiting in
  the background spacetimes of Schwarzschild and Kerr black holes. The
  initial models have a constant specific angular momentum and are
  built with a non-zero toroidal magnetic field component, for which
  equilibrium configurations have recently been obtained. In this work
  we extend our previous investigations which dealt with purely
  hydrodynamical thick discs, and study the dynamics of magnetized
  tori subject to perturbations which, for the values of the magnetic
  field strength considered here, trigger quasi-periodic oscillations
  lasting for tens of orbital periods. Overall, we have found that the
  dynamics of the magnetized tori analyzed is very similar to that
  found in the corresponding unmagnetized models. The spectral
  distribution of the eigenfrequencies of oscillation shows the
  presence of a fundamental $p$ mode and of a series of overtones in a
  harmonic ratio $2:3:\dots$. These simulations, therefore, extend the
  validity of the model of~\citet{rymz:03} for explaining the
  high-frequency QPOs observed in the spectra of LMXBs containing a
  black-hole candidate also to the case of magnetized discs with
  purely toroidal magnetic field distribution. If sufficiently compact
  and massive, these oscillations can also lead to the emission of
  intense gravitational radiation which is potentially detectable for
  sources within the Galaxy.
\end{abstract}


\date{Accepted 0000 00 00.
      Received 0000 00 00.}

%=======================================================
\section{Introduction}
\label{intro}
%=======================================================

In a series of recent papers~\citep{zanotti:03,ryz:03,zanotti:05} it
has been shown that upon the introduction of perturbations, stable
relativistic tori (or thick accretion discs) manifest a long-term
oscillatory behaviour lasting for tens of orbital periods. When the
average disc density is close to nuclear matter density, the
associated changes in the mass-quadrupole moment make these objects
promising sources of high-frequency, detectable gravitational
radiation for ground-based interferometers and advanced resonant bar
detectors, particularly for Galactic systems.  This situation applies
to astrophysical thick accretion discs formed following binary neutron
star coalescence or the gravitational core collapse of a sufficiently
massive star. If the discs are instead composed of low-density
material stripped from the secondary star in low-mass X-ray binaries
(LMXBs), their oscillations could help explaining the high-frequency
quasi-periodic oscillations (QPOs) observed in the spectra of X-ray
binaries. Indeed, such QPOs can be explained in terms of $p$-mode
oscillations of a small-size torus orbiting around a stellar-mass
black hole~\citep{rymz:03}.

The studies reported in the papers mentioned above have considered
both Schwarzschild and Kerr black holes as well as constant and
nonconstant (power-law) distributions of the specific angular momentum
of the discs. However, they have so far been limited to purely
hydrodynamical matter models, neglecting a fundamental aspect of such
objects, namely the existence of magnetic fields. There is general
agreement that magnetic fields are bound to play an important role in
the dynamics of accretion discs orbiting around black holes. They can
be the source of viscous processes within the disc through
MHD-turbulence~\citep{shakura:73}, as confirmed by the presence of the
so called magnetorotational instability (MRI) (\citet{balbus:03}) that
regulates the accretion process by transferring angular momentum
outwards.  In addition, the formation and collimation of the strong
relativistic outflows or jets routinely observed in a variety of
scales in astrophysics (from micro-quasars to radio-galaxies and
quasars) is closely linked to the presence of magnetic fields.

General relativistic magnetohydrodynamic (GRMHD hereafter) numerical
simulations provide the best approach for the investigation of the
dynamics of relativistic, magnetized accretion discs under generic
nonlinear conditions. In recent years there have been important
breakthroughs and a sustained level of activity in the modelling of
such systems, as formulations of the GRMHD equations in forms suitable
for numerical work have become available. This has been naturally
followed by their implementation in state-of-the-art numerical codes
developed by a number of groups (\textit{e.g.},~\citet{devilliers1};
\citet{harm}; \citet{komissarov}; \citet{duez:05}; \citet{shibata:05};
\citet{anninos:05}; \cite{fragile:05}; \citet{anton:06};
\citet{mckinney:06}; \citet{mizuno:06}; \citet{gr:07}) many of which
have been applied to the investigation of issues such as the MRI in
accretion discs and jet formation. Moreover, very
recently~\citet{komissarov:06} has derived an analytic solution for an
axisymmetric, stationary torus with constant distribution of specific
angular momentum and a toroidal magnetic field configuration that
generalizes to the relativistic regime a previous Newtonian solution
found by \citet{okada:89}.  Such equilibrium solution can be used not
only as a test for GRMHD codes in strong gravity, but also as initial
data for numerical studies of the dynamics of magnetized tori when
subject to small perturbations. The latter is, indeed, the main
purpose of the present paper.

In this way we aim at investigating if and how the dynamics of such
objects changes when the influence of a toroidal magnetic field is
taken into account. We discuss the implications of our findings on the
QPOs observed in LMXBs with a black hole candidate, assessing the
validity of the model proposed by~\citet{rymz:03} in a more general
context.

The paper is organized as follows: in Section 2 we briefly review the
equilibrium solution found by~\citet{komissarov:06} for a stationary
torus with a toroidal magnetic field orbiting around a black hole. The
mathematical framework we use for the formulation of the GRMHD
equations and for their implementation in our numerical code is
discussed in Section 3, while in Section 4 we describe the approach we
follow for the numerical solution of the GRMHD equations. Section 5 is
devoted to the discussion of the initial models considered, with the
results being presented in Section 6. Finally, Section 7 summarizes
the paper and our main findings. We adopt a geometrized system of
units extended to electromagnetic quantities by setting
$G=c=\epsilon_0=1$, where $\epsilon_0$ is the vacuum
permittivity. Greek indices run from 0 to 3 and Latin indices from 1
to 3.
  
%=======================================================
\section{Stationary fluid configurations with a toroidal magnetic field}
\label{II}
%=======================================================

The initial configurations we consider can be considered as the MHD
extensions to of the stationary hydrodynamical solutions of thick
discs orbiting around a black hole described by \citet{kozlowski:78},
\citet{abramowicz:78} and are built using the analytic solution
suggested recently by~\citet{komissarov:06}.

The basic equations that are solved to construct such initial models
are the continuity equation $\nabla_\mu(\rho u^\mu)=0$ for the
rest-mass density $\rho$, the conservation of energy-momentum
$\nabla_\mu T^{\mu\nu}=0$, and Maxwell's equation
$\nabla_\mu(^{*}F^{\mu\nu})=0$, where the operator $\nabla_\mu$ is the
covariant derivative with respect to the spacetime four-metric and
$^{*}F^{\mu\nu}$ is the dual of the Faraday tensor defined as
\begin{eqnarray}
^{*}F^{\mu\nu}=u^\mu b^\nu- u^\nu b^\mu.
\end{eqnarray}
In this expression $u^\mu$ is the fluid four-velocity and $b^\mu$ is
the magnetic field measured by an observer comoving with the fluid. As
usual in ideal relativistic MHD (\textit{i.e.}, for a plasma having
infinite conductivity), the stress-energy tensor $T^{\mu\nu}$ is
expressed as
\begin{eqnarray}
\label{stress-tensor}
T^{\mu\nu}\equiv (\rho h+b^{2})u^\mu u^\nu+\left(p+\frac{b^{2}}{2}\right) g^{\mu\nu} - b^\mu b^\nu,
\end{eqnarray}
\noindent
where $g^{\mu\nu}$ are the metric coefficients, $p$ is the (thermal)
pressure, $h$ the specific enthalpy, and $b^2 \equiv b^{\mu}b_{\mu}$.

The equilibrium equations are then solved to build stationary and
axisymmetric fluid configurations with a toroidal magnetic field
distribution in the tori and a constant distribution of the specific
angular momentum in the equatorial plane. The main difference of our
solution with that of \citet{komissarov:06} is that we employ a
polytropic equation of state (EOS) of the form $p=\kappa \rho^\Gamma$
for the fluid, where $\kappa$ is the polytropic constant and $\Gamma$
is the adiabatic index. Such an EOS has a well-defined physical
meaning and differs from the one used by \citet{komissarov:06},
$p=K\omega^q$, where $\omega$ is the fluid enthalpy, and $K$ and $q$
are constants.

By imposing the condition of axisymmetry and stationarity in a
spherical coordinate system (\textit{i.e.,} $\partial_{\phi} =
\partial_t =0$), the hydrostatic equilibrium conditions in the $r$ and
$\theta$ directions are given by
\begin{eqnarray}
\label{bernoulli}
 \nabla_i \ln(u_t) - \frac{\Omega \nabla_i \ell}{1- \Omega \ell} +
	\frac{\nabla_i p}{w} + \frac{\nabla_i({\cal L}b^{2})}{2{\cal L}w}=0,
\end{eqnarray} 
with $i=r, \theta$ and ${\cal L}(r,\theta) \equiv
g_{t\phi}g_{t\phi}-g_{\phi\phi}g_{tt}$. The angular velocity appearing
in (\ref{bernoulli}) is defined as
\begin{eqnarray}
\label{omega}
\Omega \equiv \frac{u^{\phi}}{u^{t}},
\end{eqnarray} 
the specific angular momentum is given by 
\begin{eqnarray}
\label{ell}
\ell \equiv -\frac{u_{\phi}}{u_{t}} \;, 
\end{eqnarray} 
and the components of the magnetic field are
\begin{eqnarray}
b^{\phi}&=&\sqrt{2p_{m}/(g_{\phi\phi}+2\ell_{0}g_{t\phi}+\ell_{0}^{2}g_{tt})},
\label{eq:bphi1}
\\ b^{t}&=&\ell_{0}b^{\phi}\;.
\label{eq:bphi}
\end{eqnarray}

Following~\citet{komissarov:06}, we consider the following EOS for the
magnetic pressure $p_{m}=M{\cal L}^{q-1} w^q$, where $M$ and $q$ are
constants, and which essentially amounts to confining the magnetic
field to the interior of the torus. Using this relation, we can
integrate eq.~(\ref{bernoulli}), which in the case of constant
specific angular momentum yields
\begin{equation}
W-W_{\rm in} + \ln\left(1+\frac{\Gamma K}{\Gamma-1}\rho^{\Gamma-1}\right) 
+ \frac{q}{q-1}M\left({\cal L} w\right)^{q-1} = 0,
\label{eq:motion1}
\end{equation}
where the potential $W$ is defined as $W \equiv \ln |u_{t}|$. Note
that in general there will be two radial locations at which $\ell_{0}$
equals the Keplerian specific angular momentum. The innermost of these
radii represents the location of the ``cusp'' of the torus, while the
outermost the ``centre''. When a magnetic field is present, the
position of the centre does not necessarily correspond with that of
the pressure maximum, as in the purely hydrodynamical
case. 

In order to solve Eq.~(\ref{bernoulli}) a number of parameters are
needed to define the initial model, namely $\kappa$, $\Gamma$, $q$,
$\ell_{0}$, $W_{\rm in}$ and the ratio of the magnetic-to-gas pressure
at the centre of the torus, $\beta_{\rm c}=(p_{m}/p)_{\rm c}$. Thus,
using the definition of $\beta_{\rm c}$, we obtain the rest-mass
density at the centre of the torus from the following expression:
\begin{eqnarray}
W_{\rm c}-W_{\rm in} &+& \ln\left(1+\frac{\Gamma}{\Gamma-1}k\rho_{\rm
  c}^{\Gamma-1}\right) \nonumber\\ &+& \frac{\beta_{\rm
    c}\Gamma}{(\Gamma-1)}\left[\frac{1}{{1}/{k\rho_{\rm
        c}^{\Gamma-1}}+{\Gamma}/({\Gamma-1})}\right]= 0.
\label{eq:motion2}
\end{eqnarray}

Finally, the equilibrium equation (\ref{bernoulli}) can be solved to
obtain the distribution of all relevant magnetohydrodynamic quantities
inside the torus (\citet{komissarov:06}).

%*************************************************************
\section{General Relativistic MHD  equations}
\label{III}
%*************************************************************

As mentioned in the Introduction, there has been intense work in
recent years on formulations of the GRMHD equations suitable for
numerical approaches~\citep{harm,devilliers1, komissarov, duez:05,
  shibata:05, anninos:05, anton:06, gr:07}. We here follow the
approach laid out in~\citet{anton:06} and adopt the $3+1$ formulation
of general relativity in which the 4-dimensional spacetime is foliated
into a set of non-intersecting spacelike hypersurfaces. The $3+1$ line
element of the metric then reads
\begin{eqnarray}
 ds^2 = -(\alpha^{2} -\beta _{i}\beta ^{i}) dx^0 dx^0 +
 2 \beta _{i} dx^{i} dx^0 +\gamma_{ij} dx^{i} dx^{j},
\label{metric}
\end{eqnarray}
where $\gamma_{ij}$ is the 3--metric induced on each spacelike slice,
and $\alpha$ and $\beta^i$ are the so-called lapse function and shift
vector, respectively.

Under the ideal MHD condition, Maxwell's equations $\nabla_\nu
^*F^{\mu\nu}=0$ reduce to the divergence-free condition for the
magnetic field
\begin{eqnarray}
\frac{\partial (\sqrt{\gamma} B^i)}{\partial x^i} &=&0,
\label{divfree} 
\end{eqnarray}
together with the induction equation for the evolution of the
magnetic field
\begin{eqnarray}
\frac{1}{\sqrt{\gamma}}\frac{\partial}{\partial t} (\sqrt{\gamma}
B^i)&=&\frac{1}{\sqrt{\gamma}} \frac{\partial}{\partial
x^j}\{\sqrt{\gamma} 
[\alpha \tilde{v}^i B^j - \alpha \tilde{v}^j B^i]\},
\label{eq:evB}
\end{eqnarray}
where $\gamma \equiv \det({\gamma_{ij}})$ and $\tilde{v}^i =
v^{i}-{\beta^i}/{\alpha}$, with $v^i$ and $B^i$ being respectively the
spatial components of the velocity and of the magnetic field, as
measured by the Eulerian observer associated to the $3+1$ splitting.

Following~\cite{anton:06}, the conservation equations for the
energy-momentum tensor given by Eq.~(\ref{stress-tensor}) together
with the continuity equation and the induction equation for the
magnetic field can be written as a first-order, flux-conservative,
hyperbolic system. The state vector and the vector of fluxes of the
fundamental GRMHD system of equations read
\begin{eqnarray}
\frac{1}{\sqrt{-g}} \left(
\frac {\partial \sqrt{\gamma}{\bf F}^{0}}
{\partial x^{0}} +
\frac {\partial \sqrt{-g}{\bf F}^{i}}
{\partial x^{i}} \right)
  = {\bf S},
\label{eq:fundsystem}
\end{eqnarray}
where $g \equiv \det(g_{\mu\nu})=\alpha\sqrt{\gamma}$. The state
vector ${\bf F}^0$ is given by
\begin{eqnarray}
{\bf F}^0 & \equiv & \left[\begin{array}{c}
D \\
S_j \\
\tau \\
B^k
\end{array}\right],
\label{state_vector}
\end{eqnarray}
with the definitions
\begin{eqnarray}
\label{conv_1}
D &\equiv& \rho W, \\    
\label{conv_2}
S_j &\equiv&\rho (h+b^{2}/\rho) W^2 v_j - \alpha b^0 b_j,  \\
\label{conv_3}
\tau &\equiv&\rho (h+b^{2}/\rho) W^2-(p+b^2/2) - \alpha^2(b^0)^2 - D,
\end{eqnarray}
and where $W$ is the Lorentz factor of the fluid. The ``fluxes'' ${\bf
  F}^i$ in eqs. (\ref{eq:fundsystem}) have instead explicit components
given by
\begin{eqnarray}
{\mathbf F}^i &\equiv& \left[\begin{array}{c}
D \tilde{v}^i\\
S_j \tilde{v}^i + (p+b^2/2) \delta^i_j - b_j B^i/W \\
\tau \tilde{v}^i + (p+b^2/2) v^i - \alpha b^0 B^i/W \\
\tilde{v}^i B^k-\tilde{v}^k B^i
\end{array}\right],
\label{flux2}
\end{eqnarray}
while the ``source'' terms ${\bf S}$ are 
\begin{eqnarray}
{\mathbf S} &\equiv& \left[\begin{array}{c}
0 \\
T^{\mu \nu} \left(
{\partial g_{\nu j}}/{\partial x^{\mu}} -
\Gamma^{\delta}_{\nu \mu} g_{\delta j} \right) \\
\alpha  \left(T^{\mu 0} {\partial {\rm ln} \alpha}/{\partial x^{\mu}} -
T^{\mu \nu} \Gamma^0_{\nu \mu} \right) \\
0^k
\end{array}\right],
\end{eqnarray}
where $0^k \equiv(0,0,0)^T$, and $\Gamma^{\mu}_{\nu\delta}$ are the
Christoffel symbols for either a Schwarzschild or Kerr black-hole
spacetime. Note that the following fundamental relations hold between
the four components of the magnetic field in the comoving frame,
$b^\mu$, and the three vector components $B^i$ measured by the
Eulerian observer
\begin{eqnarray}
\label{b0}
b^0 &=& \frac{WB^iv_i}{\alpha}, \\
\label{bi}
b^i &=& \frac{B^i + \alpha b^0 u^i}{W} \ .
\end{eqnarray}
Finally, the modulus of the magnetic field can be written as
\begin{eqnarray}
  b^2 = \frac{B^2 + \alpha^2 (b^0)^2}{W^2} \ ,
\end{eqnarray}
where $B^2 \equiv B^iB_i$.

Casting the system of evolution equations in flux-conservative,
hyperbolic form allows us to take advantage of high-resolution
shock-capturing (HRSC) methods for their numerical solution. The
hyperbolic structure of those equations and the associated spectral
decomposition of the flux-vector Jacobians, needed for their numerical
solution with Riemann solvers, is given in~\cite{anton:06}.

\begin{table*}
\caption{\label{tab1} From left to right the columns report the name
  of the model, the spin of the black hole, $a$, the specific angular
  momentum, $\ell_0$, the polytropic constant, $\kappa$, the inner and
  outer radius of the torus, $r_{\rm in}$ and $r_{\rm out}$, the
  orbital period at the point of maximum rest-mass density, $t_{\rm
    orb}$, the maximum rest-mass density, $\rho_{\rm max}$, the
  magnetic-to-gas pressure at the maximum of the rest-mass density,
  $\beta_{\rm c}$, and the maximum magnetic field, $B_{\rm max}$. For
  all models the torus-to-hole mass ratio $M_{\rm t}/M$ is 0.1 ($M=2.5
  M_{\odot}$) and the adiabatic exponent of the EOS is $4/3$.}
%\begin{center}
\begin{tabular}{cccccccccc}
\hline
Model   & $a$  & $ \ell_{0} $ 
&$\kappa $ (cgs) & $r_{\rm in}$  & $r_{\rm out}$ & $t_{\rm orb}$ (ms) &
$\rho_{\rm max}$ (cgs) &  $\beta_{\rm c}$&  $B_{\rm max}$ (G)\\

\hline  

$S1$ & 0.0  & 3.80   &   9.33${\times} 10^{13}$ & 4.57 & 15.88 & 1.86   & 
	1.25${\times} 10^{13}$  & 0.00 & 0.0\\

$S2$ & 0.0  & 3.80   &   9.21${\times} 10^{13}$ & 4.57 & 15.88 & 1.86   & 
	1.26${\times} 10^{13}$  & 0.01 & 2.50${\times} 10^{15}$ \\

$S3$ & 0.0  & 3.80   &   9.10${\times} 10^{13}$ & 4.57 & 15.88 & 1.86   & 
	1.27${\times} 10^{13}$  & 0.02 & 3.52${\times} 10^{15}$\\

$S4$ & 0.0  & 3.80   &   8.90${\times} 10^{13}$ & 4.57 & 15.88 & 1.86   & 
	1.28${\times} 10^{13}$  & 0.04 & 4.94${\times} 10^{15}$\\

$S5$ & 0.0  & 3.80   &   8.40${\times} 10^{13}$ & 4.57 & 15.88 & 1.86   & 
	1.29${\times} 10^{13}$  & 0.10 & 7.58${\times} 10^{15}$ \\

$S6$ & 0.0  & 3.80   &   7.60${\times} 10^{13}$ & 4.57 & 15.88 & 1.86   & 
	1.34${\times} 10^{13}$  & 0.20 & 1.04${\times} 10^{16}$\\

$S7$ & 0.0  & 3.80   &   6.00${\times} 10^{13}$ & 4.57 & 15.88 & 1.86   & 
	1.39${\times} 10^{13}$  & 0.50 & 1.50${\times} 10^{16}$\\

$S8$ & 0.0  & 3.80   &   4.49${\times} 10^{13}$ & 4.57 & 15.88 & 1.86   & 
	1.40${\times} 10^{13}$  & 1.00 & 1.85${\times} 10^{16}$\\

\hline

$K1$ & 0.5 & 3.30  &   2.20${\times} 10^{14}$  & 3.16 & 15.65 & 1.22   &
1.44${\times}10^{13}$  & 0.01 & 4.29${\times}10^{15}$\\	
                      
$K2$ & 0.7  & 3.00   & 2.25${\times} 10^{14}$   & 2.57 & 12.07 & 0.88   & 
2.74${\times} 10^{13}$	& 0.01 & 6.69${\times} 10^{15}$\\
	
$K3$ & 0.9  &  2.60  & 7.80${\times} 10^{14}$  &  1.77  & 19.25 & 0.56 &
1.87${\times}10^{13}$  & 0.01 & 1.15${\times}10^{16}$      \\	

\end{tabular}
%\end{center}
\end{table*}

%*************************************************************
\section{Numerical approach}
\label{IV}
%*************************************************************

The numerical code used for the simulations reported in this paper is
an extended version of the code presented in~\citet{zanotti:03,
  zanotti:05} to account for solution of the GRMHD equations. The
accuracy of the code has been recently assessed in~\citet{anton:06},
with a number of tests including magnetized shock tubes and accretion
onto Schwarzschild and Kerr black holes. The system of GRMHD equations
(\ref{eq:fundsystem}) is solved using a conservative HRSC scheme based
on the HLLE solver, except for the induction equation for which we use
the constraint transport method designed by \citet{evans:88} and
\citet{ryu:98}.  Second-order accuracy in both space and time is
achieved by adopting a piecewise-linear cell reconstruction procedure
and a second-order, conservative Runge-Kutta scheme, respectively.

The code makes use of polar spherical coordinates in the two spatial
dimensions $(r,\,\theta)$ and the computational grid consists of $N_r
\times N_{\theta}$ zones in the radial and angular directions,
respectively. The innermost zone of the radial grid is placed at
$r_{\rm {min}}=r_{\rm {horizon}}+ 0.1 $, and the outer boundary in the
radial direction is at a distance about $30\%$ larger than the outer
radius of the torus, $r_{\rm {out}}$. The radial grid has typically
$N_r\simeq 300$ and is built by joining smoothly a first patch which
extends from $r_{\rm {min}}$ to the outer radius of the torus and is
logarithmically spaced (with a maximum radial resolution at the
innermost grid zone, $\Delta r/M=1\times 10^{-3}$, where $M$ is the
mass of the black hole) and a second patch with a uniform grid and
which extends up to $r_{\rm {max}}$. On the other hand, the angular
grid consists of $N_{\theta}=100$ equally spaced zones and covers the
domain from $0$ to $\pi$.

As in the hydrodynamical code, a low density atmosphere is introduced
in those parts of the computational domain not occupied by the
torus. This is set to follow the spherically-symmetric accreting
solution described by \citet{michel:72} in the case that the
background metric is that of a Schwarzschild black hole and a modified
solution, which accounts for the rotation of the black hole
(\citet{zanotti:05}), when we consider the Kerr background
metric. Since this atmosphere is evolved as the rest of the fluid and
is essentially stationary but close to the torus, it is sufficient to
ensure that its dynamics does not affect that of the torus. This is
the case if the maximum density of the atmosphere is $5$-$6$ orders of
magnitude smaller than the central density of the torus. Note that
since we limit our analysis to isoentropic evolutions of isoentropic
initial models, the energy equation needs not to be solved. Finally,
the boundary conditions adopted are the same as those used by
\citet{font:02}.

%=======================================================
\section{Initial models}
\label{V}
%=======================================================

The initial models consist of a number of magnetized relativistic tori
which fill their outermost closed equipotential surface, so that their
inner radii coincide with the position of the cusp, $r_{\rm in}=r_{\rm
  cusp}$. In practice, we determine the positions of the cusp and of
the maximum rest-mass density in the torus by imposing that the
specific angular momentum at these two points coincides with the
Keplerian value. Clearly, different values of specific angular
momentum will produce tori with different positions of the cusp and of
the maximum rest-mass density. In a purely hydrodynamical context, the
effect on the dynamics of the tori of the distribution of specific
angular momentum, being either constant or satisfying a power-law with
$r$, was studied by~\citet{zanotti:03,zanotti:05}. In this paper,
however, we consider only magnetized tori with constant specific
angular momentum as we want to first focus on the influence a magnetic
field has on the dynamics, both in a Schwarzschild and in a Kerr
background metric.  In this way we can conveniently exploit the
analytic solution reviewed in Sect. 2 and which cannot be extended
simply to include the case of nonconstant specific angular momentum
distributions.
 
\begin{figure*}
\vskip 0.5 cm
\includegraphics[width=8.cm,angle=0]{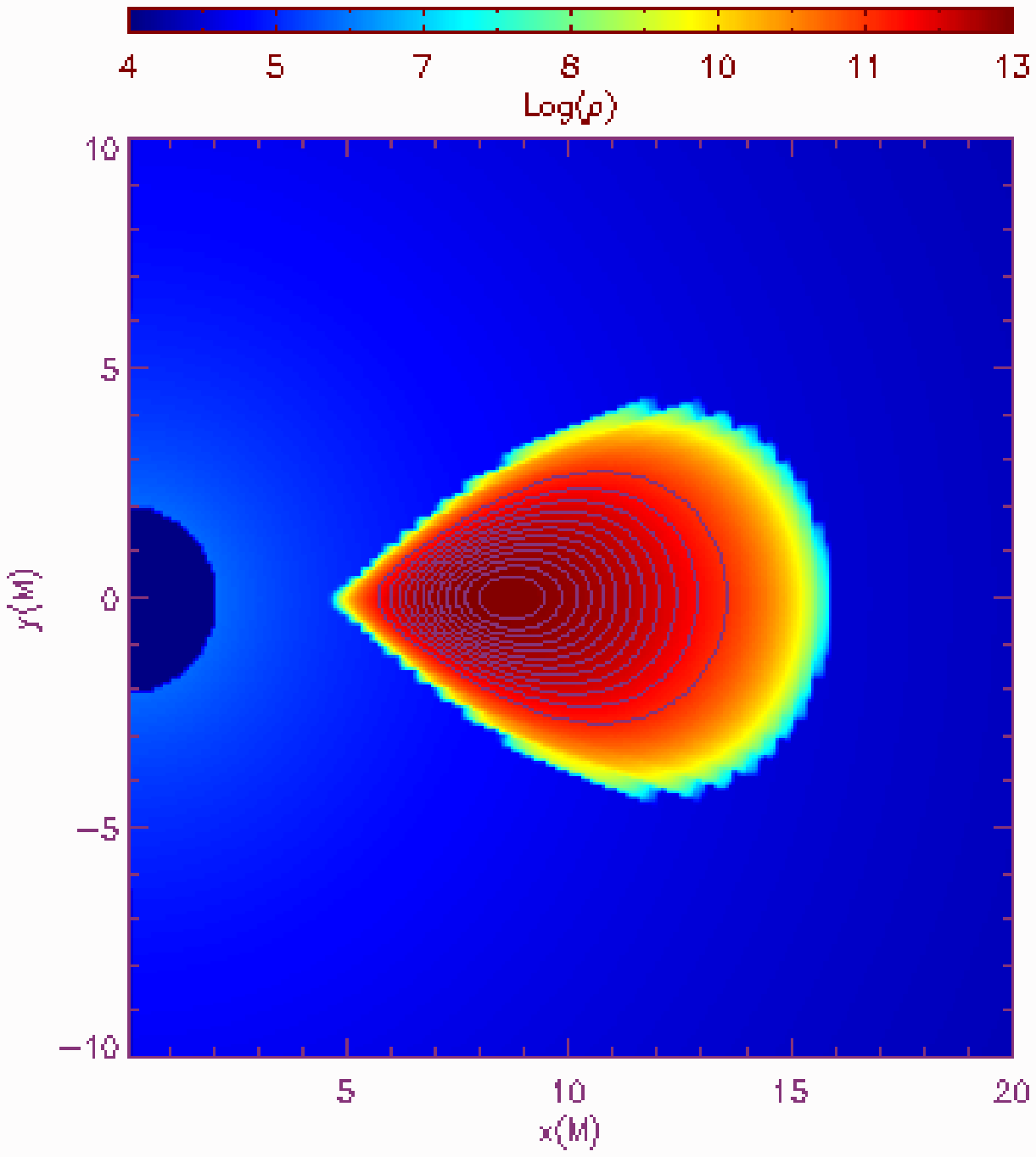}
\includegraphics[width=8.cm,angle=0]{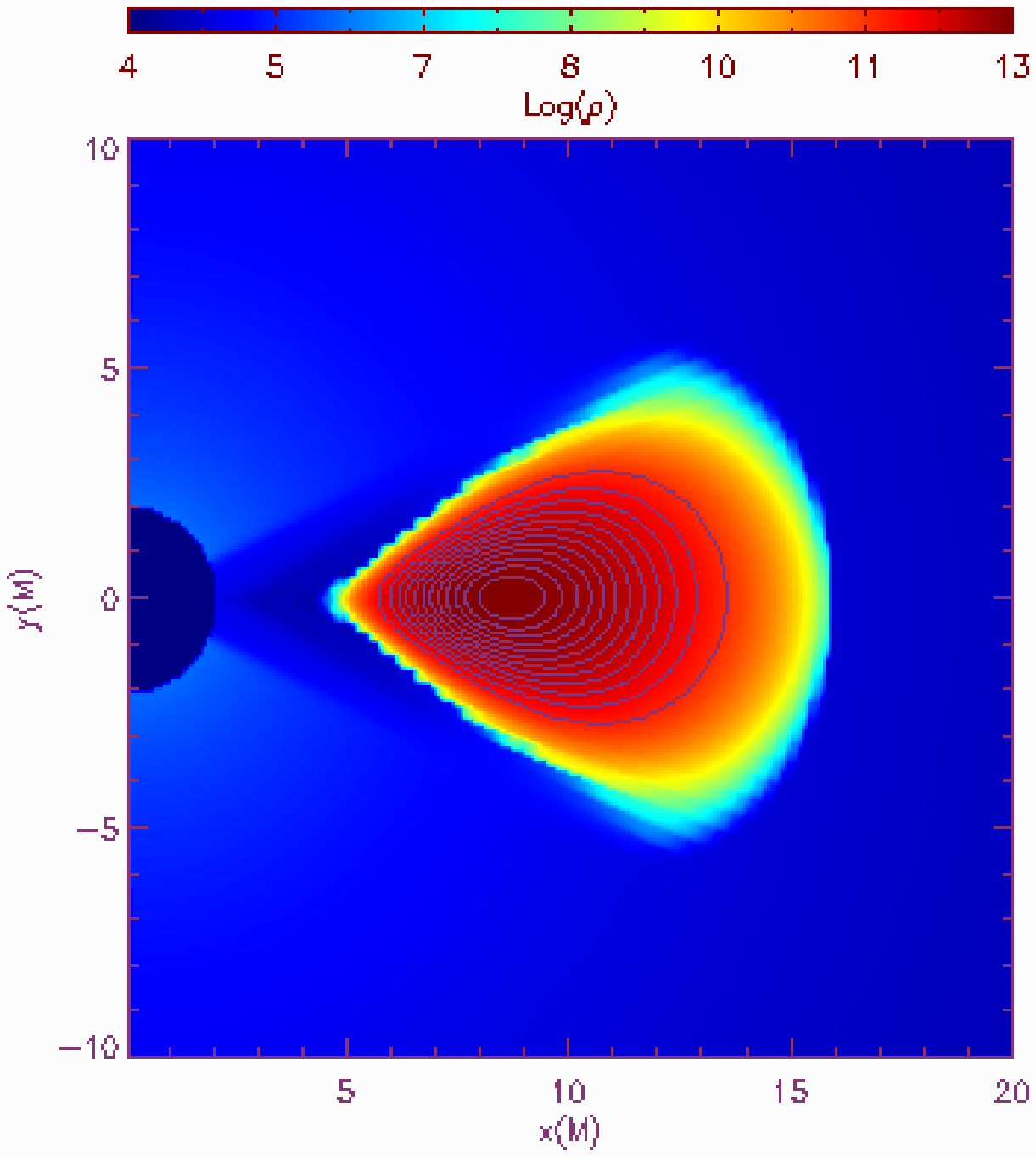}
\caption{Isocontours of the logarithm of the rest-mass density for the
  unperturbed model $S2$. The left panel shows the configuration at the
  initial time and the right panel the corresponding distribution
  after $100$ orbital timescales. The equilibrium solution is preserved
  to high accuracy.}
\label{fig0}
\end{figure*}

Once the specific angular momentum is fixed, the inner edge of the
torus $r_{\rm in}$ is determined by the potential gap at such inner
edge, $\Delta W_{\rm in}=W_{\rm in}-W_{\rm cusp}$ which, in the case
of constant specific angular momentum distributions, is defined as
\begin{eqnarray}
\label{potentilagap}
\Delta W_{\rm in}=\ln[(-u_{t})_{\rm in}] - \ln[(-u_{t})_{\rm cusp}],
\end{eqnarray}
with $\Delta W_{\rm in}=0$ corresponding to a torus filling its
outermost equipotential surface.

All of the models are built with an adiabatic index $\Gamma=4/3$ to
mimic a degenerate relativistic electron gas, and the polytropic
constant $\kappa$ is fixed such that the torus-to-black hole mass
ratio, $M_{\mathrm t}/M$, is roughly 0.1. Since the mass of the torus
is at most 10$\%$ of that of the black hole, we can neglect the
self-gravity of the torus and study the dynamics of such objects in a
fixed background spacetime (test-fluid approximation). Moreover, the
disc-to-hole mass ratio adopted here is in agreement with the one
obtained in simulations of unequal mass binary neutron star mergers
performed by~\citet{shibata:03} and \citet{shibata:05}.

Overall, we have investigated a number of different models for tori
orbiting either nonrotating or rotating black holes. In the case of
Schwarzschild black holes, the main difference among the models is the
strength of the toroidal magnetic field, which is parametrized by the
ratio of the magnetic-to-gas pressure at the centre of the disc,
$\beta_{\rm c} \equiv b^2/(2p)$. In the case of Kerr black holes, on
the other hand, we report results for tori orbiting around black holes
with spins $a=0.5, 0.7$ and 0.9, while keeping constant the
magnetic-to-gas pressure ratio at $\beta_{\rm c}=0.01$. A summary of
all the models considered is given in Table~\ref{tab1}.  

The set of models chosen here will serve a double purpose. Being 
tori with large average densities, they can provide accurate estimates
for the gravitational-wave emission triggered by the oscillations.  On
the other hand, since the ratio among the eigenfrequencies is the
astrophysically most relevant quantity and this does not depend on the
density, this set of models is also useful for analysing the
oscillation properties of the accretion discs in LMXBs.  It is also
important to note that for tori with $\beta_c>1$ the initial solution
degrades over time as a significant mass is accreted in these cases,
with an accretion rate that increases with the strength of the
magnetic field. The dependence of the stability of thick discs with
the strength of the toroidal magnetic field will be the subject of an
accompanying paper (\citet{ryzmf:07}).

The maximum strength of the magnetic field at the centre, determined
by the parameter $\beta_{\rm c}$, can be calculated through
Eq. (\ref{eq:bphi1}), which also reflects the dependence of the
toroidal magnetic field component on the background metric. The
initial models considered are such that $\beta_{\rm c}$ takes values
between $0$ and $1$, as shown in Table~\ref{tab1}. This also fixes the
overall strength of the magnetic field, whose maximum values are
reported in the tenth column of the same table. The values of the
magnetic field strength at the centre for the case of tori around a
Schwarzschild black hole range from the magnetized model $S2$ with
$2.50{\times} 10^{15}$ Gauss to model $S8$ with $1.85{\times} 10^{16}$
Gauss. These values are in good agreement with the typical values
expected to be present in the astrophysical scenarios that could form
a relativistic thick torus, such as the magnetized core
collapse~\citep{cerda:06, obergaulinger:06, shibata:06} and values
considered for the collapse of magnetized hypermassive neutron stars
by Duez et al. (2006).
     
In order to trigger the oscillations, we perturb the models reported
in Table~\ref{tab1} by adding a small radial velocity (we recall that
in equilibrium all velocity components but the azimuthal one are
zero). As in our previous work~\citep{zanotti:03,zanotti:05}, this
perturbation is parametrized in terms of a dimensionless coefficient
$\eta$ of the spherically symmetric accretion flow on to a black
hole~\citep{michel:72}, \textit{i.e.}, $v_r=\eta(v_r)_{\rm
  Michel}$. In all the simulations reported we choose $\eta=0.1$, but
the results are not sensitive to this choice as long as the
oscillations are in a linear regime (\textit{i.e.}, for $\eta \lesssim
0.2$ \citet{zanotti:03}).

Finally, it is worth commenting on the choice made for the magnetic
field distribution. On the one hand this choice is motivated by the
mere convenience of having an analytic equilibrium solution upon which
a perturbation can be introduced. On the other hand, there exists an
additional motivation which is more astrophysically motivated. As it
has been shown in recent simulations of magnetized core
collapse~\citep{cerda:06, obergaulinger:06, shibata:06} the magnetic
field distribution in the nascent, magnetized, proto-neutron stars has
a dominant toroidal component, quite irrespective of the initial
configuration. Since gravitational core collapse is one of the
processes through which thick accretion discs may form, the toroidal
initial configuration of our simulations is well justified. This
choice, however, also has an important consequence. Because of the
absence of an initial poloidal magnetic field, in fact, the
magneto-rotational instability (MRI), which could change even
significantly the dynamics of our tori, cannot develop in our
simulations. Indeed, ~\citet{fragile:05} has investigated the
oscillation of an accretion torus having an initial poloidal magnetic
field component. Although preliminary, his results suggest that the
development of the MRI and of the Papaloizou-Pringle instability
(\citet{papaloizou:84}) may damp significantly the oscillation modes
of accretion tori with poloidal magnetic fields. We will address this
question in a future work.
  
%%%%%%%%%%%%%%%%%%%%%%%%%%%
\section{Results}
\label{VI}
%%%%%%%%%%%%%%%%%%%%%%%%%%%

%%%%%%%%%%%%%%%%%%%%%%%%%%%%%%%%%%%
\subsection{Oscillation properties}
\label{VIa}
%%%%%%%%%%%%%%%%%%%%%%%%%%%%%%%%%%%

%%%%%%%%%%%%%%%%%%%%%%%
\subsubsection{Dynamics of magnetized tori}
%%%%%%%%%%%%%%%%%%%%%%%

\begin{figure*}
\includegraphics[width=3in, angle=0]{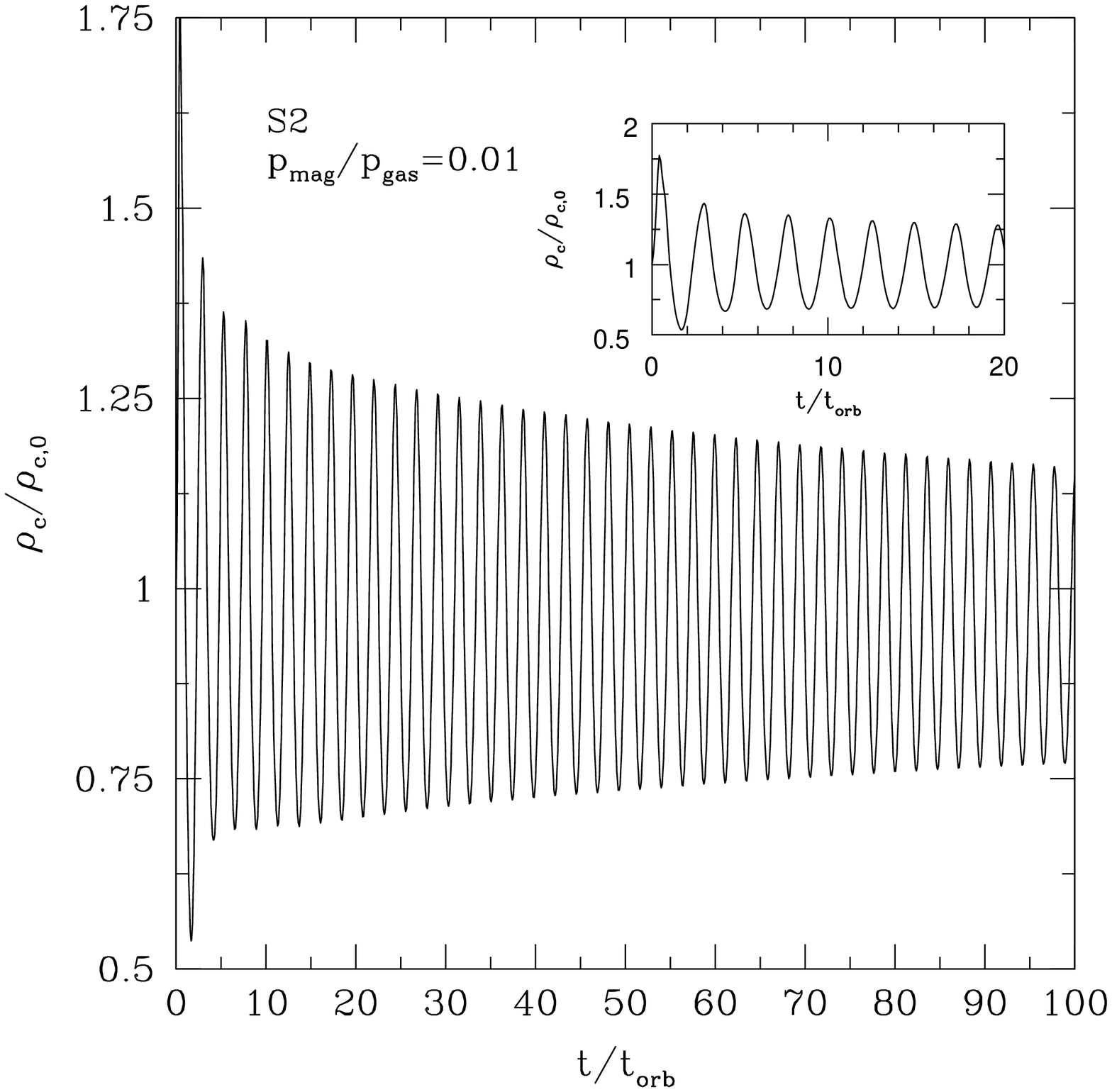}
\includegraphics[width=3in,angle=0]{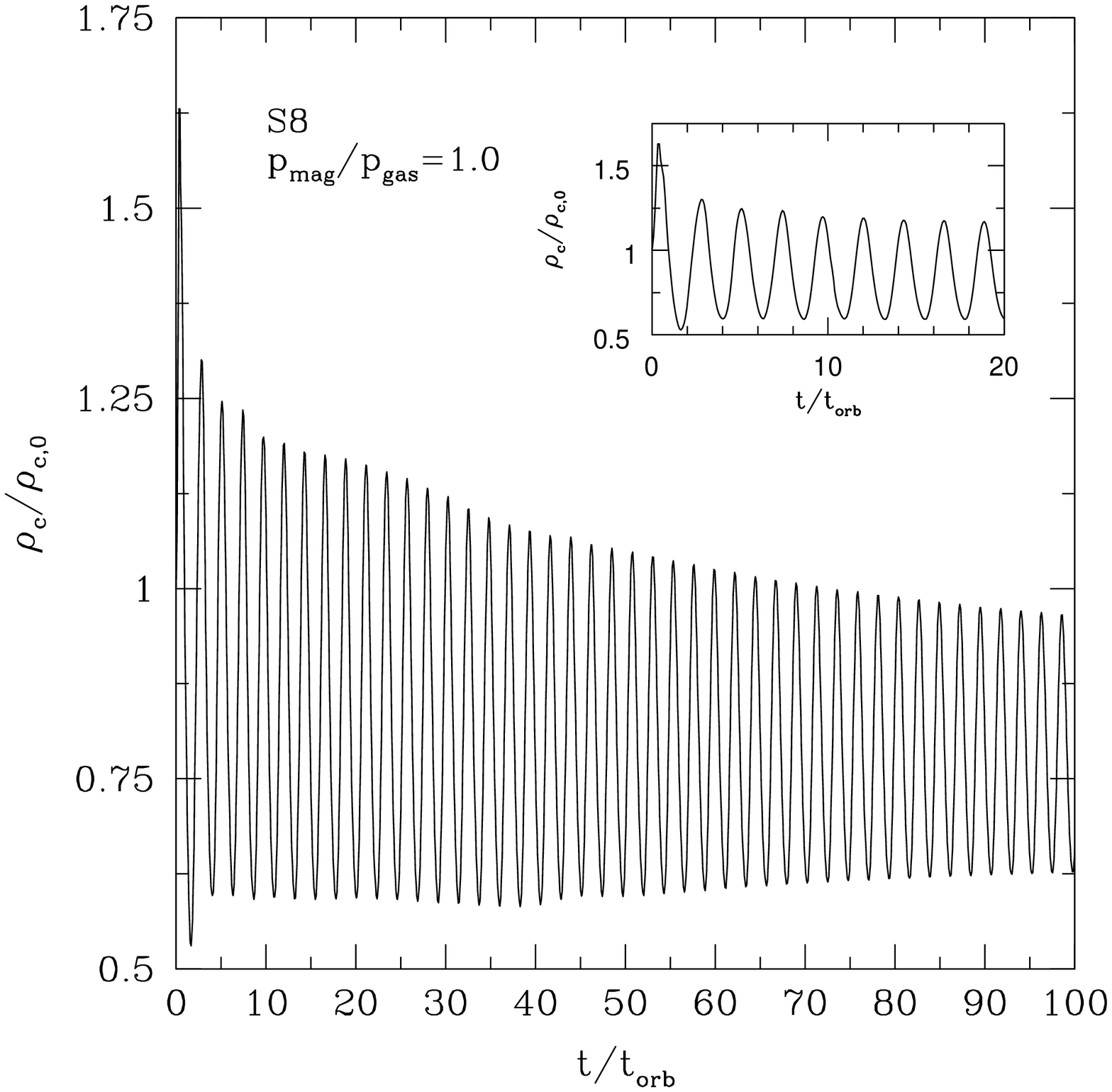}
%\vspace{-2cm}
\caption{\label{fig1} Time evolution of the central rest-mass density
  normalized to its initial value for models $S2$ (left panel) and $S8$
  (right panel). Distinctive oscillations are visible during the whole
  evolution, $t=100\,t_{\mathrm{orb}}$.}
\end{figure*}

We have first investigated equilibrium configurations of magnetized
tori by performing numerical evolutions of unperturbed tori ($\eta=0$)
and by checking the stationarity of the solution over a timescale
which is a couple of orders of magnitude larger that the dynamical
one. As representative example, we show in Fig.~\ref{fig0} the
isocontours of the logarithm of the rest-mass density of model $S2$ as
computed at the initial time $t=0$ (left panel) and at the time when
the simulation was stopped (right panel). This corresponds to
$t=100\,t_{\mathrm{orb}}$, where $t_{\mathrm{orb}}$ is the Keplerian
orbital time for a particle in a circular orbit at the centre of the
torus. Aside from the minute accretion of matter from the cusp towards
the black hole (see below), the final snapshot of the rest-mass
distribution clearly shows the stationarity of the equilibrium initial
solution. More precisely, the central rest-mass density, after a short
initial transient phase, settles down to a stationary value which
differs after 100 orbital timescales only of 2$\%$ from the initial
one. This provides a strong evidence of the ability of the code to
keep the torus in equilibrium for evolutions much longer than the
characteristic dynamical timescales of these objects.

On the left panel of Fig.~\ref{fig1} we show instead the evolution
over $100\,t_{\mathrm{orb}}$ of the central rest-mass density of the
least magnetized model $S2$, when a perturbation with parameter
$\eta=0.1$ is added to the equilibrium model\footnote{We have here
  chosen to show the evolution of the rest-mass density as this is has
  a simple physical interpretation, but all of the MHD variables
  exhibit the same harmonic behaviour.}. It is interesting to note
that despite the presence of a rather strong toroidal magnetic field,
the persistent oscillatory behaviour found in these simulations is
very similar to the one found in purely hydrodynamical
tori~\citep{zanotti:03,zanotti:05}. Note also that the small secular
decrease in the oscillation amplitude is not to be related to
numerical or physical dissipation, since the code is essentially
inviscid and the EOS used is isoentropic. Rather, we believe it to be
the result of the small but nonzero mass spilled through the cusp at
each oscillation (see also discussion below). Furthermore, on a
smaller timescale than the one shown in Fig.~\ref{fig1}, the
oscillations show a remarkable harmonic behaviour and this is
highlighted in the small insets in Fig.~\ref{fig1}. This is in stark
contrast with the results of~\citet{fragile:05}, which were obtained
with comparable numerical resolutions, but with an initial poloidal
magnetic field configuration.  In that case, in fact, the oscillations
were rapidly damped in only a few orbital periods.

\begin{figure}
\includegraphics[width=3in, angle=0]{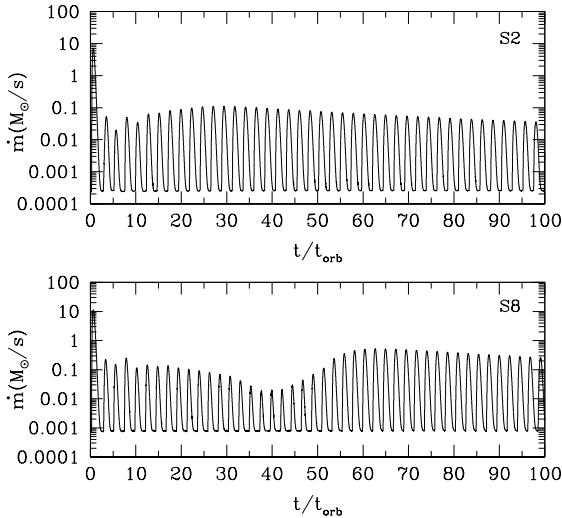}
\caption{\label{fig3} Time evolution of the mass accretion rate for
  models $S2$ and $S8$.}
\end{figure}

Results from a representative model with a higher magnetic field are
shown in the right panel of Fig.~\ref{fig1}, which again reports the
evolution of the normalized central rest-mass density for model
$S8$. Note that despite this model has a magnetic-to-gas pressure
ratio at the centre $\beta_c=1$, and hence a central magnetic field of
$\sim 2 \times 10^{16}$ G, its overall the dynamics is very similar to
that of model $S2$. Also in this case, in fact, the oscillations are
persistent during the entire evolution ($100\,t_{\mathrm{orb}}$) and
show almost no damping. However, the amplitude does show variations
over time and, most importantly, it no longer maintains a symmetric
behaviour between maxima and minima, as a result, we believe, of the
increased mass accretion through the cusp. We recall, in fact, that
all the initial models considered in our sample correspond to
marginally stable tori, \textit{i.e.}, tori filling entirely their
outermost closed equipotential surface. Any perturbation, however
small, will induce some matter to leave the equipotential surface
through the cusp, leading to the accretion of mass and angular
momentum onto the black hole. Evidence in favour of this is shown in
Fig.~\ref{fig3}, which reports the accretion mass-flux for model $S2$
(upper panel) and $S8$ (lower panel). While both reflect the
oscillations in the dynamics, they also have different mean values,
with the one relative to model $S8$ being almost an order of magnitude
larger. Note also the correlation between the fluctuations in the
mass-accretion rate and the changes in the oscillation amplitudes
shown in Fig.~\ref{fig1}. In particular, the sudden change in the
mass-flux of model $S8$ at $t\sim 35\,t_{\mathrm{orb}}$ and which
corresponds to a change in the amplitude modulation in the right panel
of Fig.~\ref{fig1}.

\begin{figure}
\includegraphics[width=3in, angle=0]{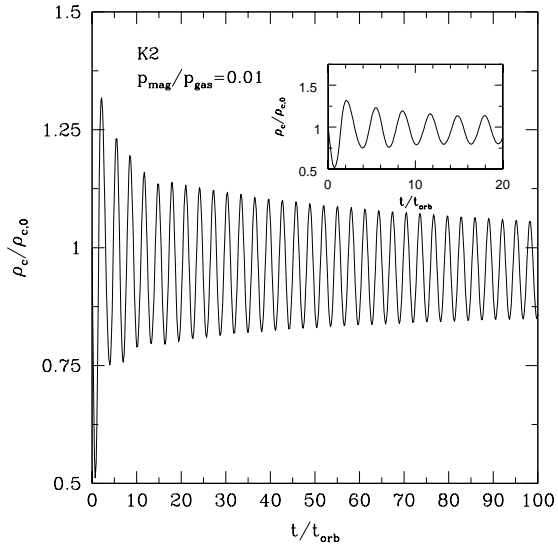}
\caption{ \label{fig4} Time evolution of the central rest-mass density
  normalized to its initial value for the model $K2$}
\end{figure}

Although the accretion-rates are well above the Eddington limit (which
is $\sim 10^{-16}\ M_{\odot}/{\mathrm s}$ for a $2.5\;M_{\odot}$ black
hole), the amounts of mass accreted by the black hole at
$t=100\,t_{\mathrm{orb}}$ is only 1.3$\%$ and 3.3$\%$ of the initial
mass for models $S2$ and $S8$, respectively. Similarly, the total
amount of angular momentum accreted at the end of the simulation would
introduce a change in the black-hole's spin of less than $1\%$ for
both models $S2$ and $S8$. Overall, therefore, these changes in the
mass and spin of the black holes are extremely small and thus justify
the use of a fixed background spacetime.  Finally, in Fig.~\ref{fig4}
we show the evolution of the normalized central rest-mass density for
model $K2$, which corresponds to a torus orbiting around a Kerr black
hole with spin $a=0.7$. Again, a perturbation with parameter
$\eta=0.1$ was added to the equilibrium model so as to investigate the
oscillatory behaviour of the torus around its equilibrium position. As
in the purely hydrodynamical case, the qualitative behaviour in models
around a Kerr black hole is very similar to that found for models
around a Schwarzschild black hole, and the dynamics shows, also in
this case, a negligible damping of the oscillations after the initial
transient.

%%%%%%%%%%%%%%%%%
\subsubsection{Power Spectra}
%%%%%%%%%%%%%%%%%

An important feature of axisymmetric $p$-mode oscillations of
accretion tori is that the lowest-order eigenfrequencies appear in the
harmonic sequence $2:3$. This feature was first discovered in the
purely hydrodynamical numerical simulations of~\citet{zanotti:03},
subsequently confirmed through a perturbative analysis in a
Schwarzschild spacetime by ~\citet{ryz:03}, and later extended to a
Kerr spacetime and to more general distributions of the specific
angular momentum by~\citet{zanotti:05}
and~\citet{montero:04}. Overall, it was found that the $2:3$ harmonic
sequence was present with a variance of $\sim 10\%$ for tori with a
constant distribution of specific angular momentum and with a variance
of $\sim 20\%$ for tori with a power-law distribution of specific
angular momentum. Since the $2:3$ harmonic sequence is the result of
global modes of oscillation, it depends on a number of different
elements that contribute to small deviations from an exact relation
among integers. The latter, in fact, should be expected only for a
perfect one-dimensional cavity, trapping the $p$ modes without
losses. In practice, however, factors such as the vertical size of the
tori, the black hole spin, the distribution of specific angular
momentum, the EOS considered, and the presence of a small but nonzero
mass-loss, can all influence this departure.

While the understanding of the properties of these modes of
oscillation has grown considerably over the last few years
(see~\citet{montero:04} for a list of references), and an exhaustive
analysis has been made in the case of relativistic slender
tori~(\cite{baf:06}), it was not obvious whether such a harmonic
sequence would still be present in the case of magnetized discs with
toroidal magnetic fields. To address this question, we have performed
a Fourier analysis of the time evolution of some representative
variables and obtained quantitative information on the quasi-periodic
behavior of the tori. In particular, for all of the models considered,
we have Fourier-transformed the evolution of the $\rm {L_2}$ norm of
the rest-mass density, defined as $\vert\vert\rho\vert\vert_2
\equiv\sum_{i=1}^{N_r} \sum_{j=1}^{N_\theta}(\rho_{ij})^2$ and studied
the properties of resulting power spectra. These, we recall, show
distinctive peaks at the frequencies that can be identified with the
quasi-normal modes of oscillation of the disc\footnote{Note that
  because of the underlining axisymmetry of our calculations, we
  cannot compute the effect of transverse hydromagnetic waves, such as
  Alfv\`en waves, propagating along the toroidal magnetic field
  lines.}. Clearly, the accuracy in calculating these eigenfrequencies
depends linearly on the length of the timeseries and is of $0.01
\mathrm{\ kHz}$ for the evolutions carried out here and that extend
for $100\,t_{\mathrm{orb}} \sim 100\,$ms.

\begin{figure*}
\includegraphics[width=3in,angle=0]{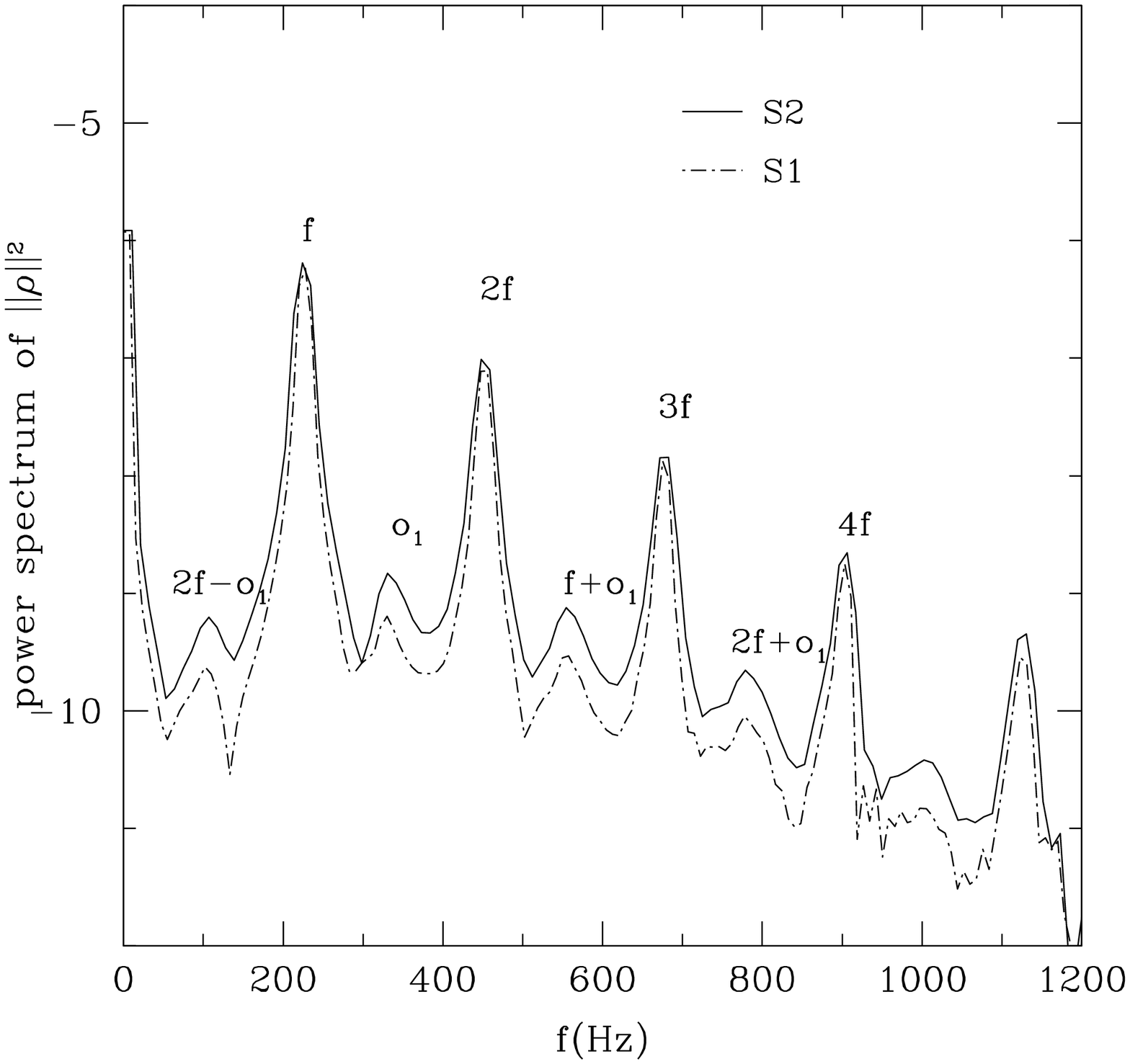}
\includegraphics[width=3in, angle=0]{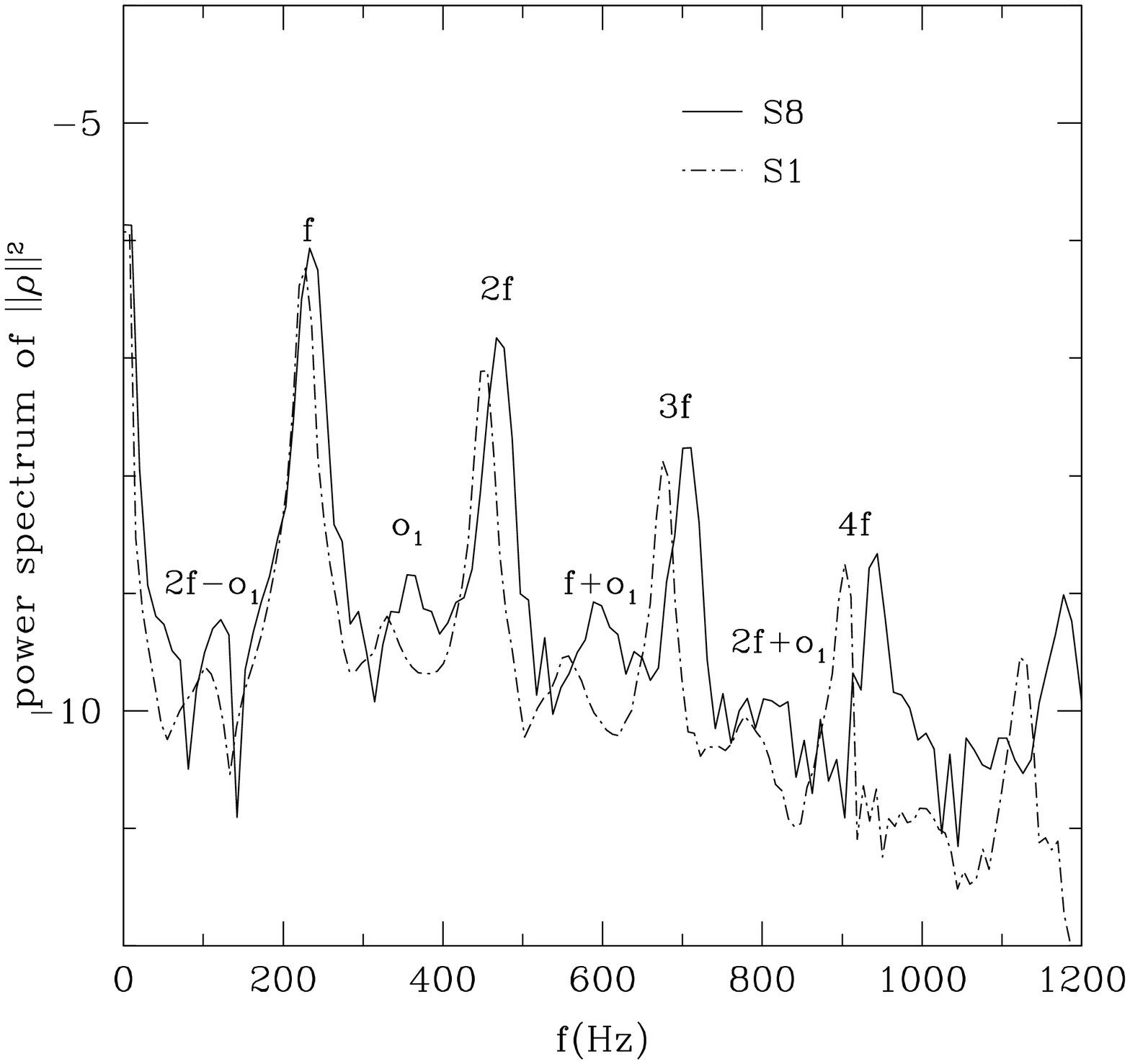}
\caption{\label{fig5} Solid lines: Power spectrum of the L-2 norm of
  the rest-mass density for models $S2$ (left panel) and $S8$ (right
  panel). The dashed lines show the corresponding PSD for the
  unmagnetized versions of both models. The units in the vertical axis
  are arbitrary and the PSDs were obtained using a Hanning filter.}
\end{figure*}

In Fig. \ref{fig5} we present the power spectra (PSD) obtained from
the $\rm {L_2}$ norm of the rest-mass density for model $S2$ (left
panel) and $S8$ (right panel); in both panels the solid lines refer to
the magnetized tori, while the dashed ones to the unmagnetized
counterpart $S1$, which is shown for reference. A rapid look at the
panels in Fig.~\ref{fig5} reveals that the overall dynamics of
magnetized tori shows features which are surprisingly similar to those
found by~\citet{zanotti:03, zanotti:05} for unmagnetized accretion
tori. Namely, the spectra have a fundamental mode $f$ (which is the
magnetic equivalent of the $p$ mode discussed
in~\citet{zanotti:03,zanotti:05}) and a series of overtones, for
which, in particular, the first overtone $o_1$ can usually be
identified clearly. Interestingly, also these spectra show the $2:3$
harmonic relation between the frequencies of the fundamental mode and
its first overtones. Such a feature remains therefore unmodified and
an important signature of the oscillations properties of magnetized
tori with a toroidal magnetic field.

It is also worth noting that in the case of mildly magnetized tori,
such as model $S2$, the similarity in the PSD is rather striking and
the two spectra differ only in the relative amplitude between the
eigenfrequencies $o_1, o_2,\ldots$ and the modes which are the result
of nonlinear coupling (\textit{e.g.}, $2f-o_1, 2f,\ldots$). On the
other hand, in the case of more highly magnetized tori, such as model
$S8$, the magnetic field strength is sufficiently large to produce
variations in the eigenfrequencies, which are all shifted to higher
frequencies, with deviations, however, which become larger for higher
overtones. While not totally unexpected [a magnetic field is known to
  increase the eigenfrequencies of magnetized stars (\cite{ns:89})],
these represent the first calculations of the eigenfrequencies of
relativistic magnetized discs and, as such, anticipate analogous
perturbative studies.

As a way to quantify the differential shift of the eigenfrequencies to
larger values, we report in Table \ref{tab2} the frequencies of the
fundamental mode, of the first overtone, and their ratio for all of
the models considered. Another analogy worth noticing in the spectra
presented in Fig.~\ref{fig5}, is the presence of nonlinear couplings
among the various oscillation modes. These modes were first pointed
out by~\citet{zanotti:05} in the investigation of the dynamics of
purely hydrodynamical tori with nonconstant specific angular momentum
in Kerr spacetime, and are the consequence of the nonlinear coupling
among modes, in particular of the $f$ and $o_1$ modes.

\begin{table}
\caption{\label{tab2} From left to right, the columns report the name
  of the model, the frequency of the fundamental mode, the frequency
  of the first overtone, their ratio, and the magnetic-to-gas pressure
  ratio at the centre of the torus.}
\begin{center}
\begin{tabular}{lllll}
\hline
Model   & $f$ (Hz)  &$o_1$ (Hz) & $ o_1/f$ & $\beta_c$ \\
\hline
$S1$   & 224  & 332 & 1.48 & 0.00\\
$S2$   & 224  & 332 & 1.48 & 0.01\\
$S3$   & 228  & 336 & 1.47 & 0.02\\
$S4$   & 229  & 333 & 1.45 & 0.04\\
$S5$   & 230  & 330 & 1.43 & 0.10\\
$S6$   & 230  & 340 & 1.48 & 0.20\\
$S7$   & 233  & 345 & 1.48 & 0.50\\
$S8$   & 235  & 341 & 1.45 & 1.00\\
\hline
$K1$   & 275  & 418 & 1.52 & 0.01\\
$K2$   & 370  & 560 & 1.51 & 0.01\\
$K3$   & 255  & 404 & 1.58 & 0.01\\
\end{tabular}
\end{center}
\end{table}

\begin{figure}
\includegraphics[width=3in, angle=0]{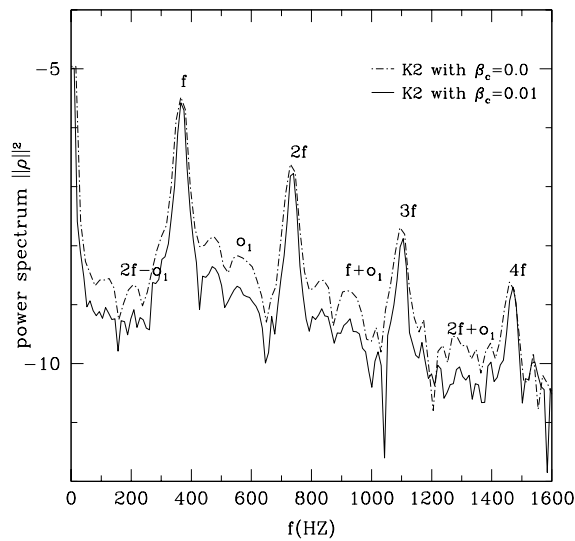}
\caption{
\label{fig6}
As Fig.~\ref{fig5} but  for model $K2$.}
\end{figure}

We complete our discussion of the spectral properties of these
oscillating discs, by showing in Fig. \ref{fig6} the PSD for model
$K2$, which, we recall, represents a torus orbiting around a Kerr
black hole with spin $a=0.7$. As for the previous spectra, the dashed
line corresponds to the unmagnetized version of model $K2$ and is
included for reference. Overall, the features observed in a Kerr
background are very similar to those found for models in the
Schwarzschild case. Also in this case, in fact, the fundamental mode,
its first overtones and the nonlinear harmonics are clearly identified
and no evidence appears of new modes related to the presence of a
toroidal magnetic field.

As a final remark we note that the $2:3$ ratio among the different $p$
modes has a relevance also in a wider context. We recall, in fact,
that among the several models proposed to explain the QPOs observed in
LMXBs containing a black hole candidate, the one suggested
by~\citet{rymz:03} is particularly simple and is based on the single
assumption that the accretion disc around the black hole terminates
with a sub-Keplerian part, {\it i.e} a torus of small size. A key
point of this model is the evidence that in these objects the
frequencies of the fundamental mode and the first overtone are in the
$2:3$ harmonic sequence in a very wide space of parameter. The
simulations presented here further increase this space, extending it
also to the case of magnetized tori and thus promoting the validity of
this model for QPOs to a more general and realistic scenario.

%%%%%%%%%%%%%%%%%%%%%%%%%%%%%%%%%%%%%%%%%%%%%%%%
\subsection{Gravitational-wave emission}
\label{VIb}
%%%%%%%%%%%%%%%%%%%%%%%%%%%%%%%%%%%%%%%%%%%%%%%%

As pointed out by~\citet{zanotti:03} the oscillating behavior of
perturbed accretion tori is responsible for significant changes of
their mass quadrupole moment. As a result, these changes determine the
emission of potentially detectable gravitational radiation if the tori
are compact and dense enough. This could be the case if the tori are
produced via binary neutron star mergers or via gravitational collapse
of the central core of massive stars. In this section, we extend the
analysis of~\cite{zanotti:03,zanotti:05} for unmagnetized discs and
investigate the gravitational-wave emission from constant angular
momentum magnetized tori orbiting around black holes.

Although more sophisticated approaches involving perturbative
techniques around black holes can be employed to study the
gravitational-wave emission from these tori (\cite{nagar:05,
  ferrari:06, nagar:07}), we here resort to the simpler and less
expensive use of the Newtonian quadrupole approximation
(\citet{zanotti:03}), which has been suitably modified to account for
the presence of a magnetic field, as done by~\citet{kotake:04}. In
particular, the quadrupole wave amplitude $A_{20}^{\rm E2} $, and
which is the second time derivative of the mass quadrupole moment, is
computed through the ``stress formula'' (\citet{obergaulinger:06})
\begin{eqnarray}
\label{stress}
&& \hskip -0.5cm
A_{20}^{\rm E2} =
	k \!\int\!\! r^2 {\rm d}r {\rm d}z \bigg[(f_{rr} (3 z^2 \!-\! 1) \!+\!
	f_{\theta\theta} (2 \!-\! 3 z^2) \!-\!
	f_{\phi\phi}  \nonumber \\
	& & \quad 
        -6 z f_{r\theta} \sqrt{1 \!-\! z^2}
	\biggl. - r \rho \frac{\partial \Phi}{\partial r}
	(3 z^2 \!-\! 1) 
        + 3 z \rho \frac{\partial \Phi}{\partial \theta}
	\sqrt{1 \!-\! z^2}\bigg]  \;, \nonumber \\
\end{eqnarray}
where $ k = 16 \pi^{3/2} / \sqrt{15} $, $z\equiv \cos\theta$, $f_{ij}
\equiv \rho v_i v_j -b_ib_j$, and $\Phi$ is the gravitational
potential, and is approximated at the second post-Newtonian order from
the metric function $g_{rr} = 1 -2\Phi + 2\Phi^{2}$.

\begin{figure}
\centering
\includegraphics[width=3in,angle=0]{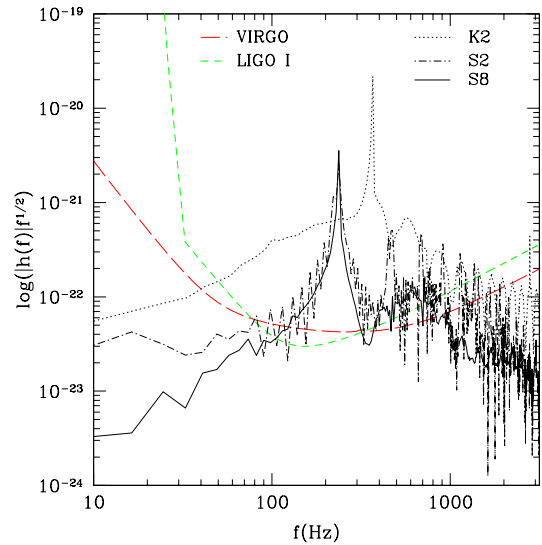}
\caption{ Comparison between the power spectrum $|h(f)|\sqrt{f}$ of
  the wave signal for models $S2$, $S8$, and $K2$ and the strain
  sensitivity of LIGO (dashed line), Virgo (long-dashed line).
\label{fig7}
 }
\end{figure}

Figure~\ref{fig7} shows a spectral comparison between the designed
strain sensitivity of the gravitational-wave detectors Virgo and LIGO,
and the logarithm of the power spectrum $|h(f)|\sqrt{f}$ of the
gravitational-wave signals for models $S2$, $S8$, and $K2$ (similar
graphs are obtained also for the other models). Note that all of the
sensitivity curves displayed in this figure assume an optimally
incident wave in position and polarization (as obtained by setting the
beam-pattern function of the detector to one), and that the sources
are assumed to be located at a distance of 10 kpc.

From Fig.~\ref{fig7} it is clear that all our models lie well above
the sensitivity curves of the detectors for galactic sources and also
that there are no significant differences in the power spectra as the
magnetic field strenght is increased. Interestingly, however, the
signal from a torus orbiting around a Kerr black hole is clearly
distinguishable from the one around a Schwarzschild black
hole. Besides having a fundamental mode at higher frequencies, in
fact, also the amplitude is about one order of magnitude larger as a
result of it being closer to the horizon and with a comparatively
larger central density. As expected from the similarities in the
dynamics, the signal-to-noise of these magnetized models is very
similar to one of the corresponding unmagnetized tori, and we refer to
\citet{zanotti:05} for a detailed discussion.

%%%%%%%%%%%
\section{Conclusions}
\label{VII}
%%%%%%%%%%%
 
We have presented and discussed the results of numerical simulations
of the dynamics of magnetized relativistic axisymmetric tori orbiting
in the background spacetime of either Schwarzschild or Kerr black
holes. The tori, which satisfy a polytropic equation of state and have
a constant distribution of the specific angular momentum, have been
built with a purely toroidal magnetic field component. The
self-gravity of the discs has been neglected and, as the models
considered are all marginally stable to accretion, the minute
accretion of mass and angular momentum through the cusp is not
sufficient to affect the background black hole metric.

The use of equilibrium solutions for magnetized tori around black
holes has allowed us to study their oscillation properties when these
are excited through the introduction of small perturbations. In
particular, by considering a representative sample of initial models
with magnetic-field strengths that ranged from $2.5\times 10^{15}$ G
up to equipartition, and GRMHD evolutions over 100 orbital periods, we
have studied the dynamics of these discs and how this is affected by a
magnetic field.

Overall, we have found the behaviour of the magnetized tori to be very
similar to the one shown by purely hydrodynamical tori
(\cite{zanotti:03,zanotti:05}). As in the hydrodynamical case, in
fact, the introduction of perturbations triggers quasi-periodic
oscillations lasting tens of orbital periods, with amplitudes that are
modified only slightly by the small loss of matter across the
cusp. The absence of an inital poloidal magnetic field has prevented
the development of the magneto-rotational instability, which could
influence the oscillation properties and thus alter our conclusions
(\cite{fragile:05}). Determining whether this is actually the case
will be the focus of a future work, where a more generic magnetic
field configuration will be considered.

As for unmagnetized tori, the spectral distribution of the
eigenfrequencies shows the presence of a fundamental $p$-mode and of a
series of overtones in a harmonic ratio $2:3:\dots$. The analogy with
purely hydrodynamical simulations extends also to the nonlinear
harmonics in the spectra and that are the consequence of the nonlinear
coupling among modes, (in particular the $f$ mode and of its first
overtone $o_1$). Also for them we have found a behaviour which is
essentially identical to that found in unmagnetized discs. In summary,
no new modes have been revealed by our simulations, and in particular
no modes which can be associated uniquely to the presence of a
magnetic field. Nevertheless, the influence of the magnetic field is
evident when considering the absolute values of the eigenfrequencies,
which are shifted differentially to higher frequencies as the strength
for the magnetic field is increased, with an overall relative change
which is $\sim 5\%$ for a magnetic field near equipartition.

Besides confirming the unmagnetized results, the persistence of the
$2:3$ ratio among the different $p$ modes also has an important
consequence. It allows, in fact, to extend to a more general and
realistic scenario the validity of the QPO model presented
by~\citet{rymz:03} and \citet{sr:06}, and which explains the QPOs
observed in the x-ray luminosity of LMXBs containing a black hole
candidate with the quasi-periodic oscillations of small tori near the
black hole. The evidence that this harmonic ratio is preserved even in
the presence of toroidal magnetic fields, provides the model with
additional robustness.

When sufficiently massive and compact, the oscillations of these tori
are responsible for an intense emission of gravitational waves and
using the Newtonian quadrupole formula, conveniently modified to
account for the magnetic terms in the stress-energy tensor, we have
computed the gravitational radiation associated with the oscillatory
behaviour. Overall, we have found that for Galactic sources these
systems could be detected as they lie well within the sensitivity
curves of ground-based gravitational-wave interferometers.
        
As a concluding remark we note that our discussion here has been
limited to tori with magnetic fields whose pressure is at most
comparable with the gas pressure, \textit{i.e.}, $\beta_c \leq 1$. The
reason behind this choice is that while in equilibrium, the magnetized
tori with a purely toroidal magnetic field are not necessarily
stable. Rather, indications coming both perturbative calculations and
from nonlinear simulations, suggest that these tori could be
dynamically unstable for sufficiently strong magnetic fields. The
results of these investigations will be presented in a forthcoming
paper~(\cite{ryzmf:07}).

%%%%%%%%%%%%%%
\section*{Acknowledgments}
%%%%%%%%%%%%%%

It is a pleasure to thank Chris Fragile and Shin Yoshida for useful
discussions and comments. Pedro Montero is a VESF fellow of the
European Gravitational Observatory (EGO-DIR-126-2005). This research
has been supported by the Spanish Ministerio de Educaci\'{o}n y
Ciencia (grant AYA2004-08067-C03-01) and through the SFB-TR7
``Gravitationswellenastronomie'' of the DFG. The computations were
performed on the computer ``CERCA2'' of the Department of Astronomy
and Astrophysics of the University of Valencia.
 
\bibliographystyle{mn2e}

\label{lastpage}  
\end{document}